\documentclass[prl,aps,twocolumn,showpacs,preprintnumbers,amsmath,amssymb]{revtex4}
\usepackage{dcolumn}
\usepackage{bm}
\usepackage{epsfig}

\begin{document}


\title{Magnetic field-tuned quantum critical point in CeAuSb$_{2}$}

\author{L. Balicas,$^1$ S. Nakatsuji,$^2$ H. Lee,$^3$ P. Schlottmann,$^1$, T. P. Murphy$^1$, and Z. Fisk$^3$}
\affiliation{$^1$National High Magnetic Field Laboratory, Florida
State University, Tallahassee-FL 32306, USA. \\
$^2$Department of Physics, University of Kyoto, Kyoto
606-8502, Japan. \\
$^3$Department of Physics, University of California Davis,
Davis, California 95616, USA.}

\date{\today}%
\begin{abstract}

Transport, magnetic and thermal properties at high magnetic fields
($H$) and low temperatures ($T$) of the heavy fermion compound
CeAuSb$_2$ are reported. At $H=0$ this layered system exhibits
antiferromagnetic order below $T_N = 6$ K. Applying $B$ along
the inter-plane direction, leads to a continuous suppression of
$T_N$ and a quantum critical point at $H_c \simeq 5.4$ T.
Although it exhibits Fermi liquid behavior within the N\'{e}el
phase, in the paramagnetic state the fluctuations associated
with $H_c$ give rise to unconventional behavior in the resistivity
(sub-linear in $T$) and to a $T \ln {T}$ dependence in the magnetic
contribution to the specific heat. For $H > H_c$ and low $T$ the
electrical resistivity exhibits an unusual $T^3$-dependence.
\end{abstract}

\maketitle

When the long-range order is suppressed to zero temperature
by tuning an external variable, such as pressure, chemical
composition or magnetic field $H$, the system is said
to exhibit a quantum critical point (QCP)\cite{QCP,stuart}.
The magnetic field is an ideal control parameter, since it
can be reversibly and continuously tuned towards the QCP.
Two compounds with field-tuned QCP, YbRh$_2$Si$_2$ and
Sr$_3$Ru$_2$O$_7$, reached prominence due to the non-Fermi
liquid (NFL) behavior triggered by the quantum fluctuations
associated with the QCP. In this letter we present a Ce-compound,
CeAuSb$_2$, exhibiting a field-tuned QCP and unusual
transport and thermodynamic properties. All three systems
have a field-tuned QCP as a common thread, yet their
behavior in high fields and low $T$ are considerably
different.

At zero field YbRh$_2$Si$_2$ exhibits a second-order phase
transition into an antiferromagnetic (AF) state at $T_N =70$
mK \cite{trovareli}. A magnetic field applied along the
inter-plane direction drives $T_N$ to zero at a critical
$H_c \simeq 0.66$ T, leading to NFL behavior, i.e., a
logarithmic increase of $C_{e}(T)/T$ and a quasilinear
$T$ dependence of the electrical resistivity $\rho$ below
10 K \cite{gegenwartprl}. Above $H_c$ Fermi liquid (FL)
behavior is recovered ($\rho\propto AT^2$ and constant
Sommerfeld coefficient $\gamma$), with $A(H)$ and
$\gamma(H)^2$ displaying a $1/(H-H_c)$ divergence as
$H \rightarrow H_c$ \cite{gegenwartnature}. A similar
trend was recently found in YbAgGe \cite{budkoprb}.

Field-tuned anomalous metallic behavior in the vicinity of
metamagnetism (MM) was studied in detail in Sr$_3$Ru$_2$O$_7$
\cite{grigera}. At a MM transition the magnetization $M$
increases rapidly over a narrow range of fields. The transition
is of first order, since there is no broken symmetry involved,
and terminates in a critical ``end" point $(H^{\star}, T^{\star})$
\cite{millisprl}. In the anisotropic MM transition of
Sr$_3$Ru$_2$O$_7$, $T^{\star}$ is found to decrease continuously
as $H$ is rotated towards the inter-plane c-axis \cite{perry},
thus opening the possibility of a QCP in a first order transition
\cite{grigera,millisprl}. This scenario is supported by the
$T$-linear dependence in $\rho$ \cite{grigera}, a divergence
of the coefficient $A$ of the resistivity \cite{grigera}, the
enhancement of the effective mass of the quasiparticles \cite{borzi},
and the $\ln{T}$-dependence of the specific heat $\gamma$ \cite{Zhou}.
Remarkably, at very low $T$ and very close to the critical field
$H_c$, $\rho$ displayed a $T^3$-dependence \cite{grigera}. The
QCP of a MM transition is also believed to cause the rich phase
diagram of URu$_2$Si$_2$ at high fields \cite{marcelo}.

Among Ce compounds, so far there is evidence for a field-tuned QCP
only in CeCoIn$_5$ \cite{bianchiandpaglioni} and CeIrIn$_5$
\cite{capan}. In these systems the QCP is believed to give rise to
a (possibly unconventional) superconducting (SC) phase, in
addition to NFL behavior. In CeCoIn$_5$ the nature of the
magnetic correlations at low $T$ is still unclear, in part due to
the close proximity to SC, while in CeIrIn$_5$ the quantum critical
behavior close to the metamagnetic (MM) transition is still under
investigation.

\begin{figure}[tbh]
\begin{center}
\epsfig{file = 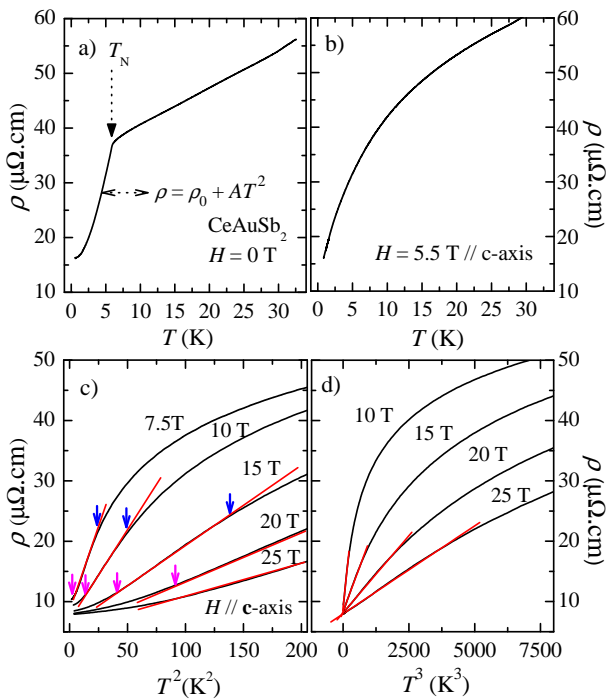, width=8 cm} \caption{a)
$T$-dependence of the in-plane resistivity $\rho$ of a CeAuSb$_2$
single crystal. At $T_N$ a pronounced change in slope is observed,
below which FL behavior is recovered. b) This behavior is
suppressed and a sub-linear $T$-dependence is observed when $T_N
\rightarrow 0$. c) At higher fields a $T^2$-dependence is obtained
but only over a limited range of $T$ as indicated by the vertical
arrows (red lines are linear fits to a $T^2$-dependence). d) At
the lowest $T$ and largest $H$ a $T^3$-dependence is observed (red
lines are linear fits to a $T^3$-dependence) }
\end{center}
\end{figure}

In this letter we report on the anomalous properties of CeAuSb$_2$.
CeAuSb$_2$ is a tetragonal metallic compound, which at $H=0$
orders AF \cite{onuki} with $T_N = 6.0$ K \cite{comment}. For
$T < T_N$, $\rho(T)$ has the $A T^2$ dependence typical of a FL
and the extrapolation of $C_e/T$ to $T=0$ yields a Sommerfeld
coefficient of $\gamma \sim 0.1$ J/mol.K$^2$. Hence, CeAuSb$_2$
can be considered a system of relatively light heavy-fermions.
Above $T_N$, on the other hand, $\rho(T)$ displays a $T^{\alpha}$
dependence with $\alpha \lesssim 1$ and, $C_e/T$ has a $-\ln T$
dependence, both characteristic of NFL behavior due to a nearby
QCP. A magnetic field along the inter-plane direction leads to
two subsequent metamagnetic transitions and the concomitant
\emph{continuous} suppression of $T_N$ to $T=0$ at $H_c = 5.3
\pm 0.2$ T. As the AF phase boundary is approached from the
paramagnetic (PM) phase, $\gamma$ is enhanced and the $A$
coefficient of the resistivity diverges as $(H-H_c)^{-1}$.
When $T$ is lowered for $H \sim H_c$, the $T$-dependence of
$\rho$ is sub-linear and the one of $C_e/T$ is approximately
$-\ln T$. These observations suggest the existence of a
field-induced QCP at $H_c$. At higher fields an unconventional
$T^3$-dependence emerges in $\rho$ and becomes more prominent
as $H$ increases, suggesting that the FL state is \emph{not}
recovered for $H \gg H_c$.

\begin{figure}[htb]
\begin{center}
\epsfig{file = 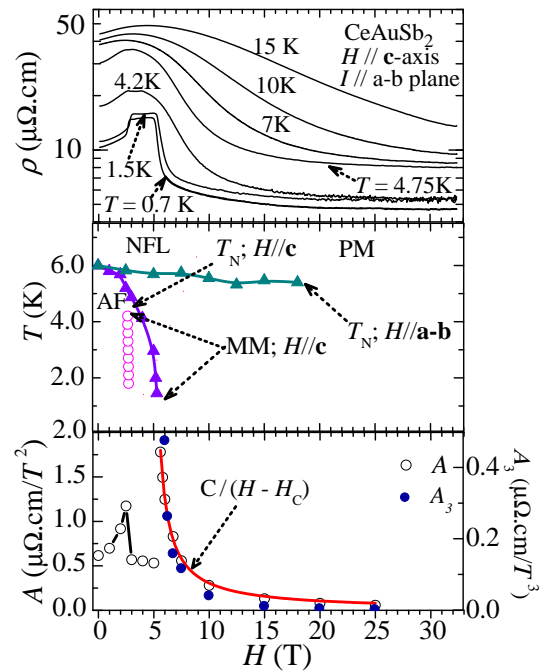, width=7 cm} \caption{Upper panel:
In-plane resistivity $\rho$ of CeAuSb$_2$ as a function of $H$.
Middle panel: The resulting phase diagram indicating the AF phase
boundary, the PM region, and the region in field where NFL
behavior is found. Arrows indicate the critical fields of the
metamagnetic transitions. Note that the effect of $H$ on the
N\'{e}el temperature $T_N$ is very anisotropic. Lower panel:
Coefficients $A$ (open circles) and $A_3$ (blue circles) of the
resistivity as a function of $H$. See text for details.}
\end{center}
\end{figure}

Single crystals of CeAuSb$_2$ were grown by the self-flux method,
as described in Ref. \cite{canfield}, using high purity starting
materials with excess Sb as a flux. Microprobe analysis confirms
the stoichiometric composition of the samples as well as the
absence of sub-phases. Electrical transport measurements were
performed with the Lock-In technique in both, a Bitter and a
superconducting magnet, coupled to cryogenic facilities . The
magnetization as a function of $T$ and $H$ was obtained in a
Quantum Design DC SQUID, as well as with a high field vibrating
sample magnetometer. The heat capacity was measured in a Quantum
Design PPMS system using the relaxation time technique. The heat
capacity measurements were extended to higher temperatures in
order to resolve the crystalline electric field (CEF) scheme. A
Schottky-like peak centered around $T \simeq 50$ K was observed,
suggesting an excitation energy of $\Delta = 110$ K between the
ground and first excited doublets.

Figure 1a) displays the in-plane electrical resistivity $\rho$
as function of $T$ for $H=0$ T. The vertical arrow indicates
the onset of AF order, below which a $T^2$-dependence of $\rho$
is obtained. Above $T_N$ a NFL-like $T^{\alpha}$-dependence with
$\alpha \lesssim 1$ is observed. As $H$ increases the AF order
is gradually suppressed and in the vicinity of the AF to PM
phase boundary a linear dependence on $T$ emerges (see Fig. 1b)).
At higher fields FL behavior, i.e., $\rho = \rho_{0} + AT^{2}$,
is found over a limited range of $T$ as indicated in Fig. 1c)
by the vertical arrows. The red lines are least square fits to
the $T^2$-dependence. The slope, given by the $A$ coefficient,
decreases as $H$ increases. At very low $T$ and very high $H$
$\rho$ displays an anomalous $T^3$-dependence (see Fig. 1d)),
suggesting, on the one hand, an unusual scattering mechanism,
and, on the other hand, that the $T^2$-dependence may just be
a crossover regime between $T^{\alpha}$ with $\alpha < 1$ and
the $T^3$ regions.

\begin{figure}[htb]
\begin{center}
\epsfig{file = 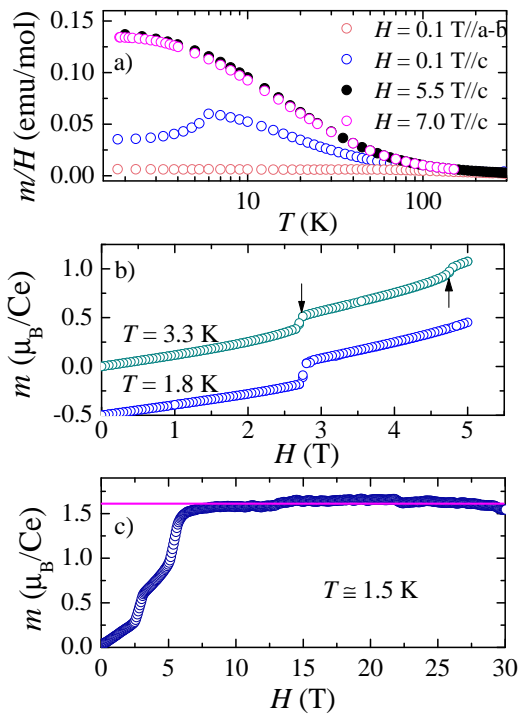, width=6.8 cm} \caption{a) Magnetic
moment $m$ of CeAuSb$_2$ divided by the field $H$, as a function
of $T$ for several values of $H$. b) $m$ of Ce, measured in a
SQUID magnetometer, as a function of field $H$ for $T=3.3$ K and
$1.8$ K (The vertical scale for the latter is off-set by $-0.5$).
Vertical arrows indicate the fields of both metamagnetic
transitions. c) $m$ as a function of $H$ at $T \simeq 1.5$ K
measured with a vibrating sample magnetometer.}
\end{center}
\end{figure}

The $H$ dependence of $\rho$ sheds more light on the role of the
magnetic fluctuations (see upper panel of Fig. 2). For $T > 6$ K,
a crossover from positive to negative magnetoresistance is seen in
$\rho(H)$ at $H \sim 5$ T. As $T$ is lowered below 6 K, two MM
transitions emerge, with negative magnetoresistance developing at
the AF to PM transition. The magnetic field suppresses the AF
spin-correlations, giving rise to a gradual alignment of the spins
and reducing this way the spin-flip scattering. The mechanism of
the anomalous $T^3$-dependence in $\rho(H)$ is believed to be
related to the alignment of the Ce-spins. As seen below, the
saturation of the magnetization $m$ suggests that the system is
nearly half-metallic for very large $H$.

The middle panel of Fig. 2 shows the phase diagram
resulting from transport and magnetization measurements.
The blue triangles indicate the boundary between the
PM and the AF phases for $H \parallel$ c-axis, while
the green ones define the phase boundary for $H$ in the
a-b plane. Notice the remarkable anisotropy. Surprisingly
no hysteresis was detected, neither in $T$ nor in $H$
scans, despite the abrupt changes in the slope of
$\rho(H)$. This suggests that the transition between
the PM and AF phase is a weak first order one or
unexpectedly of second-order. The first MM-transition at
$H_{\text{MM}} \simeq 2.8$ T, which could correspond to
a spin-flop transition, is indicated with open circles.

The field dependence of the $A$ and $A_3$ coefficients
(pre-factors of the $T^2$ and $T^3$ terms) of the
resistivity is displayed in the lower panel of Fig. 2.
Both coefficients diverge as $H \to H_c$ and the red
line is a fit to $C/(H - H_c)$. The same dependence
was reported for YbRh$_2$Si$_2$ \cite{gegenwartprl}
and interpreted as a dramatic increase of the
quasiparticles linewidth throughout the entire Fermi
surface. The $A$-coefficient also increases at the
first MM-transition due to an enhancement in the
scattering of the quasiparticles.

\begin{figure}[htb]
\begin{center}
\epsfig{file = 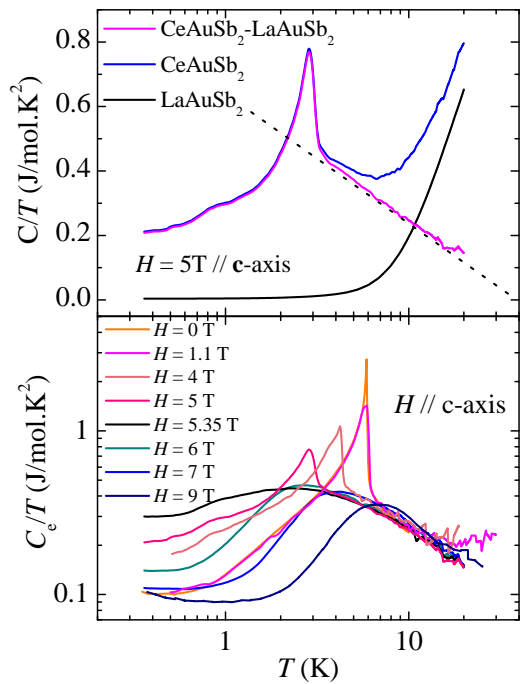, width=6.8 cm} \caption{Upper panel:
Heat capacity divided by temperature $C/T$ vs. $T$ for CeAuSb$_2$
with $H=5$ T applied along the c-axis (blue line), as well as for
LaAuSb$_2$ at $H=0$ T (black line). The difference is the magnetic
contribution to the heat capacity $C_{\text{e}}/T$ (in magenta),
which shows a $\ln{T}$-dependence. Lower panel: $C_{\text{e}}/T$
for several magnetic fields. All curves collapse into a single
curve at high $T$ and $\gamma$ ($\gamma = C_{\text{e}}/T$ as
$T\rightarrow0$) increases as $H$ approaches $H_c$.}
\end{center}
\end{figure}

The magnetic moment $m$ of a CeAuSb$_2$ single crystal divided by
the field is shown in Fig. 3a) as a function of $T$. For small
fields and higher $T$, $m/H$ follows a Curie-Weiss law, yielding
an antiferromagnetic Weiss-temperature $\theta_C = 12.25$ K and an
effective moment $p_{eff} \approx 2.26 \mu_B$, for the field
applied along the c-axis. Notice that the AF transition is
observed only for $H \parallel c$, which is the magnetic easy
axis. At higher fields, $m/H$ saturates, suggesting the complete
polarization of the Ce-spins. In Fig. 3b) both metamagnetic
transitions are seen as steps in the magnetization $m$ as a
function of $H$. Both traces contain field-up and down sweeps
measured in a SQUID magnetometer. Again, as for the resistivity,
no hysteresis is observed. The field of the first MM-transition is
only weakly $T$-dependent. In contrast, as already seen in the
resistivity data, the $T_N$ for the second MM-transition (AFM to
PM) is markedly field-dependent. Fig. 3c) shows $m$ as a function
of $H$ (up to 30 T) at $T = 1.5$ K. Above $H \sim 7$ T, $m$
saturates at $m \simeq 1.65 \mu_{B}$, a value that differs from
the free ion moment of Ce due to crystalline electric field
effects.

\begin{figure}[htb]
\begin{center}
\epsfig{file=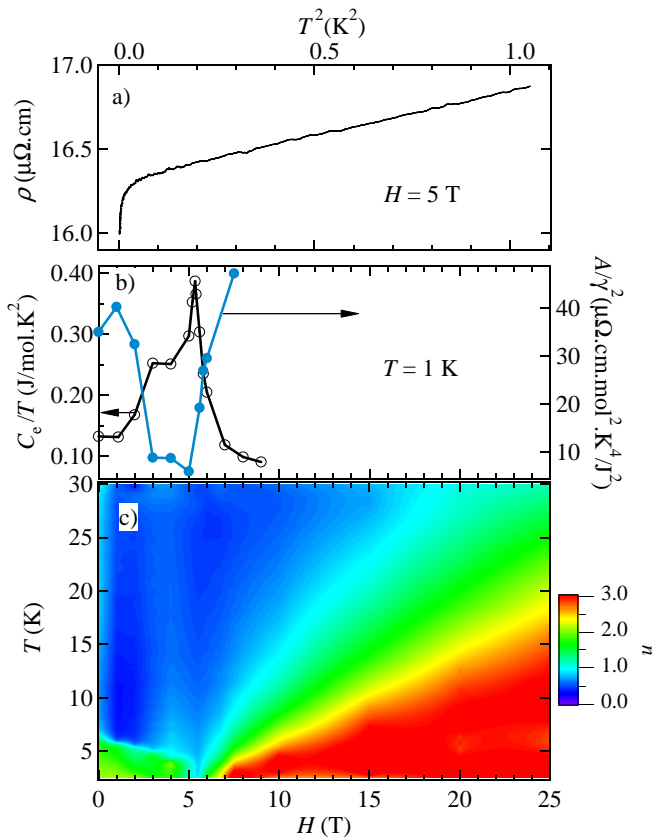, width=8.6 cm} \caption{a) The
$T^2$-dependence of $\rho$ down to $T \simeq 25$ mK for $H = 5$ T
$\lesssim H_c$. Notice the pronounced deviation to the $T^2$
dependence emerging at the lowest $T$s. b) The dependence of
$C_{\text{e}/T}$ for $T = 1$ K (black markers) and the
Kadowaki-Woods ratio $A/\gamma^2$ (blue markers) as a function of
$H$. There is a sharp increment of $C_{\text{e}}/T$ at $H_c$. c)
The exponent $n$ of $\rho(T)$ in the $T$ - $H$ plane.}
\end{center}
\end{figure}

The upper panel of Fig. 4 shows $C(T)/T$ as a function of
$T$ for CeAuSb$_2$ at $H=5$T, and for its isostructural
non-magnetic analog LaAuSb$_2$ at $H=0$T. The large
peak for CeAuSb$_2$ signals the AF transition. The
subtraction of both curves yields the magnetic
contribution to the heat capacity $C_{\text{e}}(T)/T$.
For $3 < T < 20$ K, $C_{\text{e}}(T)/T$ displays a
$-\ln(T)$ NFL-like dependence. The lower panel
of Fig. 4 shows $C_{\text{e}}(T)/T$ as a function
of $T$ for several values of $H$. In the PM phase
all curves collapse into a single trace, suggesting
that the origin of the $\ln(T)$-dependence is the
magnetic pre-critical fluctuations to the AF order.
Furthermore, the effective mass of the quasi-particles,
as given by $\gamma_{\text{e}}$ ($C_{\text{e}}(T)/T$
for $T \rightarrow 0$) increases considerably as
$H\rightarrow H_c$.

Fig. 5a) shows the $T^2$-dependence of $\rho$ down to $T
\simeq 25$ mK for $H = 5$ T $\lesssim H_c$. There is
evidence for a pronounced reduction $\rho$ at the lowest
$T$. In Fig. 5b) we display $C_{\text{e}}(T)/T$ at
$T=1$ K as a function of $H$ by the black (open) markers.
This shows that $\gamma$ dramatically increases at the
critical field, but that it remains finite. In other
words, the $-\ln(T)$-dependence does not continue to
very low $T$. The blue (closed) represent the Kadowacki-Woods
ratio, $A/\gamma^{2}$, as a function of $H$. The fact
that this ratio is not constant indicates that there
is more than one energy scale involved.

Fig. 5c) depicts a qualitative sketch of the phase diagram.
It shows the dependence of the exponent $n
\simeq \partial \ln{(\rho(T) - \rho_{0})}/ \partial \ln{T}$
on $H$ and $T$. Here different values of $\rho_0$ were used
in the PM and AF phases. The PM phase is indicated by the
blue region which is influenced by the QCP leading to the
anomalous NFL value $n \lesssim 1$. The FL state (in green)
is recovered below the N\'{e}el temperature but is gradually
suppressed as $H \rightarrow H_c$. As for YbRh$_2$Si$_2$
\cite{gegenwartnature,gegenwartprl} a FL-like $n=2$
exponent is obtained above $H_c$ but only over a limited
range of $T$. A spin-polarized metallic phase (in red)
with an anomalous $T^3$-dependence is observed for $H >
H_c$ at low $T$. Besides Sr$_3$Ru$_2$O$_7$ \cite{grigera},
such dependence has also been predicted for half-metallic
ferromagnets in terms of a one magnon scattering process
\cite{furukawa}. This could be a plausible explanation
for our system, which is spin-polarized precisely in the
region where $n=3$ is observed. Although CeAuSb$_2$ is
not a FM in zero-field, it could have similar properties
to a FM half-metal in a strong magnetic field.

In summary, electrical transport, magnetization and thermal
properties indicate that CeAuSb$_2$ displays either a weak
first-order or a second-order phase transition from a NFL-like PM
metallic phase to a FL phase with long-range AF-order upon cooling
at zero field. A magnetic field continuously reduces $T_N$, which
vanishes for a field of $H_c = 5.3 \pm 0.2$ T along the
inter-layer direction. The physical properties of the PM phase
emerging at $H_c$ are anomalous, i.e., $\rho(T) \propto T^{1/2}$,
and $C(T)/T \propto -\ln(T)$, and are a strong indication that
$H_c$ corresponds to a field-tuned QCP. In addition, the $A$ and
$A_3$ coefficients of the resistivity diverge as $H \rightarrow
H_c$. The consequences of the QCP in CeAuSb$_2$ are different and
perhaps more dramatic than for other field-tuned QCP systems. This
is the case because the FL phase is not recovered for $H$ and $T$
sufficiently far away from the QCP. The narrow region in Fig. 5
with $n=2$ should be considered a crossover region and not a FL
phase. The field-tuned QCP systems represent a challenge from the
theoretical perspective, since the different compounds have some
common aspects, but do \emph{not} seem to belong to the same
universality class.

This work is sponsored by the National Nuclear Security
Administration under the Stewardship Science Academic Alliances
program through DOE grant DE-FG03-03NA00066. This work was
performed at the NHMFL which is supported by NSF through
NSF-DMR-0084173 and the State of Florida. LB acknowledges
support from the NHMFL in-house research program and PS
through grants (DOE) DE-FG02-98ER45707 and (NSF) DMR01-05431.



\begin{thebibliography}{}


\bibitem{QCP} See, for instance, S. Sachdev, \emph{Quantum Phase Transitions} (Cambridge
Univ. Press, Cambridge, 1999).
\bibitem{stuart} For a review in materials and properties see, G. R Stewart, Rev. Mod. Phys. \textbf{73}, 797 (2001).
\bibitem{trovareli} O. Trovarelli \emph{et al.}, Phys.\ Rev.\ Lett.\ \textbf{85}, 626 (2000).
\bibitem{gegenwartprl} P. Gegenwart \emph{et al.}, Phys.\ Rev.\ Lett.\ \textbf{89}, 056402
(2002).
\bibitem{gegenwartnature} J. Custers \emph{et al.}, Nature
(London) \textbf{424}, 524 (2003).
\bibitem{budkoprb} S. L. Bud'ko \emph{et al.}, Phys.\ Rev.\ B \textbf{69}, 014415
(2004).
\bibitem{grigera}  S. A. Grigera \emph{et al.}, Science  \textbf{294}, 329 (2001).
\bibitem{millisprl} A. J. Millis \emph{et al.}, Phys.\ Rev.\ Lett.\ \textbf{88}, 217204 (2002).
\bibitem{perry} R. S. Perry \emph{et al.}, Phys.\ Rev.\ Lett.\ \textbf{92}, 166602 (2004).
\bibitem{borzi} R. A. Borzi \emph{et al.}, Phys.\ Rev.\ Lett.\ \textbf{92}, 216403 (2004).
\bibitem{Zhou} Z.X. Zhou \emph{et al}, Phys. Rev. B {\bf 69}, 140409(R) (2004).
\bibitem{marcelo} M. Jaime \emph{et al.}, Phys.\ Rev.\ Lett.\ \textbf{89}, 288101 (2002);
N. Harrison \emph{et al.}, \emph{ibid} \textbf{90}, 096402 (2003);
K. H. Kim \emph{et al.} \emph{ibid} \textbf{91}, 256401 (2003).
\bibitem{bianchiandpaglioni} A. Bianchi \emph{et al.}, Phys. Rev. Lett. \textbf{91}, 257001
(2003); J. Paglione \emph{et al.}, ibid \textbf{91}, 246405 (2003).
\bibitem{capan} C. Capan \emph{et al.}, cond-mat/0404333.
\bibitem{onuki} A. Thamizhavel \emph{et al.}, Phys.\ Rev.\ B \textbf{68}, 054427
(2003).
\bibitem{comment} The $T_N$ reported here is higher than the value reported in Ref. \cite{onuki}, due to differences in sample quality.
\bibitem{canfield} K.D. Myers \emph{et al.}, J.\ Magn.\ Magn.\ Mater.\ \textbf{205}, 27
(1999).
\bibitem{furukawa} N. Furukawa, J.\ Phys.\ Soc.\ Jap.\ \textbf{69}, 1954 (2000), and references therein.
\end{thebibliography}

\end{document}